___

# A multimodal approach for Parkinson disease analysis

**Marcos Faundez-Zanuy [1], Antonio Satue-Villar [1], Jiri Mekyska [2], Viridiana Arreola [3,4], Pilar Sanz [3], Carles Paul [1], Luis Guirao [3], Mateu Serra [3,4], Laia Rofes [4], Pere Clavé [3,4], Enric Sesa-Nogueras [1], Josep Roure [1]**

[1] Fundació Tecnocampus, Avda. Ernest Lluch 32, 08302 Mataró, SPAIN
`{faundez, satue, paul, sesa, roure}@tecnocampus.cat`
[2] Brno University of Technology, Brno, Czech Republic
[3] Hospital de Mataró, Consorci Sanitari del Maresme, SPAIN
`{msanz,mserra}@csdm.cat`
4 Centro de Investigación Biomedica en Red de Enfermedades Hepaticas y Digestivas. Barcelona, SPAIN
`{laia.rofes, pere.clave}@ciberehd.org`

**Abstract.** Parkinson's disease (PD) is the second most frequent neurodegenerative disease with prevalence among general population reaching 0.1-1 %, and an annual incidence between 1.3-2.0/10000 inhabitants. The mean age at diagnosis of PD is 55 and most patients are between 50 and 80 years old. The most obvious symptoms are movement-related; these include tremor, rigidity, slowness of movement and walking difficulties. Frequently these are the symptoms that lead to the PD diagnoses. Later, thinking and behavioral problems may arise, and other symptoms include cognitive impairment and sensory, sleep and emotional problems. In this paper we will present an ongoing project that will evaluate if voice and handwriting analysis can be reliable predictors/indicators of swallowing and balance impairments in PD. An important advantage of voice and handwritten analysis is its low intrusiveness and easy implementation in clinical practice. Thus, if a significant correlation between these simple analyses and the gold standard video-fluoroscopic analysis will imply simpler and less stressing diagnostic test for the patients as well as the use of cheaper analysis systems.

## 1 Introduction

In this study we will focus on three kinds of signals, but the first step will be focused on speech signals and dysphagia.

It is based on a collaboration be-tween an engineering faculty and a Hospital.

___

## 1.1 Voice analysis

In the PD patient, dysphagia is usually accompanied by other oro-bucal symptoms such as hypokinetic dysarthria. Some studies have reported that the presence of both symptoms usually correlates and that voice disorders could be anticipatory of swallowing impairment [1]. Other studies concluded that a clear post-swallow voice quality provides reasonable evidence that penetration-aspiration and dysphagia are absent [2]. Voice analysis is a safe, non-invasive, and reliable screening procedure for patients with dysphagia which can detect patients at high risk of clinically significant aspiration [6]. The volume-viscosity swallow test (V-VST) was developed at the Hospital de Mataró to identify clinical signs of impaired efficacy (labial seal, oral and pharyngeal residue, and piecemeal deglutition) and impaired safety of swallow (voice changes, cough and decrease in oxygen saturation $\geq 3$ %) [4]. The V-VST allows quick, safe and accurate screening for oropharyngeal dysphagia (OD) in hospitalized and independently living patients with multiple etiologies. The V-VST presents a sensitivity of 88.2 % and a specificity of 64.7 % to detect clinical signs of impaired safety of swallow (aspiration or penetration). The test takes 5-10 min to complete and is an excellent tool to screen patients for OD. It combines good psychometric properties, a detailed and easy protocol designed to protect safety of patients, and valid end points to evaluate safety and efficacy of swallowing and detect silent aspirations [3]. However, nowadays voice assessment is usually done by subjective parameters and a more exhaustive and objective evaluation is needed to understand its relationship with dysphagia and aspiration, as well as the potential relevance of voice disorders as a prognostic factor and disease severity marker. Hypokinetic dysarthria is a speech disorder usually seen in PD which affects mainly respiration, phonation, articulation and prosody. Festination is the tendency to speed up during repetitive movements. It appears with gait in order for sufferers to avoid falling down and also in handwriting and speech. Oral festination shares the same pathophysiology as gait disorders [7]. Voice analysis allows the assessment of all these parameters and has been used to evaluate the improvement of PD after treatment [8-11,17]. Voice impairments appear in early stages of the disease and may be a marker of OD even when swallow disorders are not clinically evident, which would allow to establish early measures to prevent aspiration and respiratory complications. Oropharyngeal dysphagia is a common condition in PD patients. In a recent meta-analysis, the prevalence of PD patients who perceive difficulty in swallowing was estimated at 35% but when an objective swallowing assessment was performed, the estimated prevalence of OD reached 82% [20]. This underreporting calls for a proactive clinical approach to dysphagia, particularly in light of the serious clinical consequences associated to OD in these patients. Dysphagia can produce two types of severe complications; a) alterations in the efficacy of deglutition that may cause malnutrition or dehydration which may occur in up to 24% of PD patients [21], and b) impaired safety of swallow, which may lead to aspiration pneumonia with high mortality rates (up to 50%) [22-23]. Aspiration pneumonia remains the leading cause of death among PD patients.

___________________________________________________________________

### 1.2 Balance and falls

Postural instability is one of the cardinal signs in PD. It becomes more preva-lent and worsens with disease progression and represents one of the most disabling symptoms in the advanced stages of PD, as it is associated with falls and loss of independence [12]. Balance impairments represent a major burden with high impact on individual's functional capacity, mobility, quality of life and survival. Overall, more than half of patients with PD experience falls. Falls are a major milestone in the evolution of PD because their severe consequences such as bone fractures or head injuries leading to disability, institutionalization and death [13]. Most falls occur during posture changes and are unrelated to extrinsic factors, but are dependent upon intrinsic deficits of balance control. However, pathophysiology of balance disorders and postural instability in PD is not well understood. Posturography allows an objective assessment of balance parameters and posturographic studies have contributed to significant advances in understanding the pathophysiology of postural instability in PD, but it still remains to be fully clarified, partially due to the difficulty to distinguish between the disease process and the compensatory mechanisms and also due to the lack of standardized techniques to measure balance. Dopaminergic treatments can provide improvements in postural instability in early- to mid-stage of PD but the effects tend to decrease with time consistent with spread of the disease process to non-dopaminergic pathways.

### 1.3 Handwriting

Handwriting skill degradation appears in early stages of PD so handwriting analysis is also of interest in the assessment of the disease progression. Alterations of central dopaminergic neurotransmissions adversely affect movement execution during handwriting and automatic execution of well-learned movements. Drawing exercises in a digitalized tablet allows the accurate evaluation and quantification of size, velocity, acceleration, stroke duration and other parameters of handwriting [18-19]. Although beneficial effects of dopaminergic treatments in kinematics of handwriting movements have been reported, PD patients do not reach an undisturbed level of performance, suggesting that dopamine medication results in partial restoration of automatic movement execution [14-15]. Some authors have shown altered parameters in PD as well as a recovery to the skill of a healthy person after medication with apomorphine [16]. Handwriting tests are useful for assessing the effect of medication and for determining the dosage of drugs for a specific patient.

## 2 Dysphagia and Speech analysis

This multimodal analysis has started on speech signals in the context of dysphagia test. A large number of group of people suffer dysphagia, as summarized in table 1, as well as their effect.
The medical term for any difficulty or discomfort when swallowing is dysphagia. A normal swallow takes place in four stages, and involves 25 different muscles and five different nerves. Difficulties at different stages cause different problems and symptoms. The four stages of swallowing are the following ones:

1. The sight, smell, or taste of food and drink triggers the production of saliva, so that when you put food in your mouth (usually voluntarily) there is extra fluid to make the process of chewing easier.
2. When the food is chewed enough to make a soft bolus, your tongue flips it towards the back of the mouth to the top of the tube, which leads down to your stomach. This part of your throat is called the pharynx. This part of swallowing is also voluntary.
3. Once the bolus of food reaches your pharynx, the swallowing process becomes automatic. Your voice box (the larynx) closes to prevent any food or liquid getting into the upper airways and lungs, making the food bolus ready to pass down your throat (known as the oesophagus).
4. The oesophagus, which is a tube with muscular walls that contract automatically, then propels the food down to the stomach.

**Table 1.** Groups affected by dysphagia and its sympthoms.

| Group of people | Effects |
|---|---|
| Elderly people | 45 % find some difficulty in swallowing; 65 % of those living in residential or nursing homes: chewing and swallowing muscles are weaker, loss of teeth and saliva production reduced. |
| Stroke sufferers | 40 % nerves, muscles and cognitive/brain function affected |
| Multiple sclerosis or Parkinson's disease | Nervous system and muscles affected |
| Alzheimer's disease sufferers and severe depressives | Cognitive/brain function affected |
| Motor Neuron Disease | Nervous system, nerves and muscles affected |
| People with cancer of the throat and/or mouth | Nerves and muscles damaged by disease and treatment |
| People with head and neck injuries | Nerves and muscles damaged |

Some signs of dysphagia are:
1 Swallow repeatedly.

______________________________________________________________

2    Cough and splutter frequently.
3    Voice is unusually husky and you often need to clear your throat.
4    When you try to eat you dribble. Food and saliva escape from your mouth or even your nose.
5    Find it easier to eat slowly.
6    Quite often keep old food in your mouth, particularly when you have not had a chance to get rid of it unseen.
7    Feel tired and lose weight.

## 2.1 Gold standard for dysphagia analysis

The courrent approach for dysphagia analysis has been developed by some of the medical authors of this paper, and can be summarized in figure 1.

Process for dysphagia analysis based on three liquids of different viscosity and three different volumes per liquid. After swallowing each liquid and volume a word is pronounced by the patient and a speech therapist evaluates the voice quality in a subjective way (just listening to the speech signal).

In those cases where a possible dysphagia problem exists, a videofluoroscopic analysis is perfomed. This diagnose is more invasive as it implies radiation, but it is the procedure to have physical evidence of swal-lowing problems. Figure 2 shows Videofluoroscopic pictures and oropharyngeal swallow response during the ingestion of a 5 mL nectar bolus in: (a) a healthy individual; (b) an older patient with neurogenic dysphagia and aspiration associated with stroke. An increased total duration of the swallow response may be seen, as well as a delayed closure of the laryngeal vestibule and delayed aperture of the upper sphincter. The white dot indicates the time when contrast penetrates into the laryngeal vestibule, and the red dot indicates passage into the tracheobronchial tree (aspiration). GPJ = glossopalatal junction, VPJ = velopalatal junction, LV= laryngeal vestibule, UES = upper esophageal sphincter.

The main goal of the first step of this study is to evaluate if an automatic tool based on speech analysis can be developed to support medical decision during the test depicted in figure 1.

____________________________________________________________________

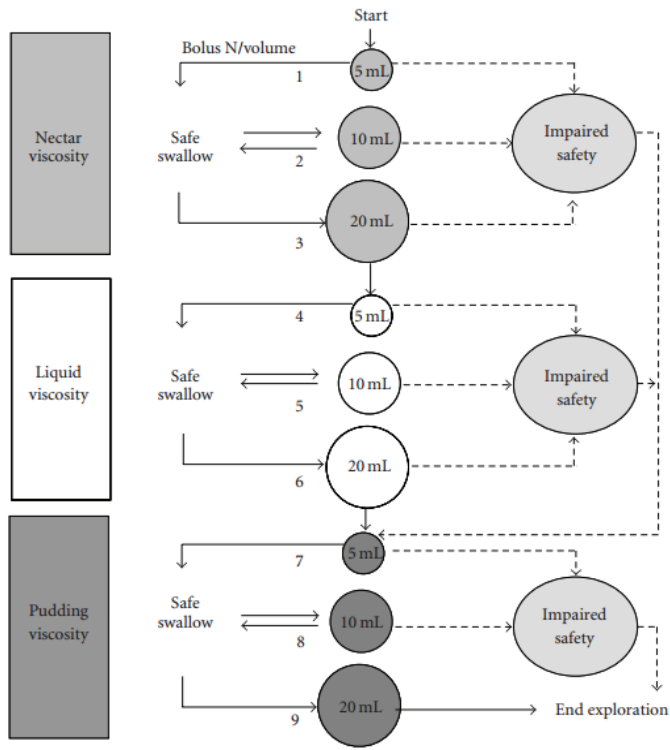

**Fig. 1.** Process for dysphagia analysis based on three liquids of different viscosity and three different volumes per liquid.

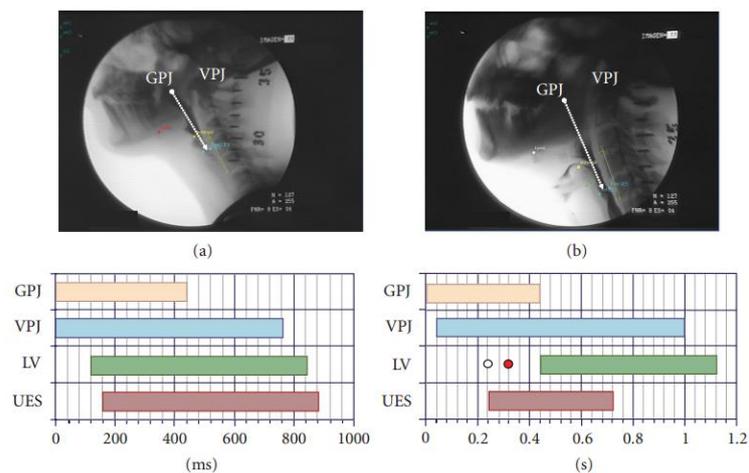

**Fig. 2.** Videofluorescence images for (a) a healthy individual; (b) an older patient with neurogenic dysphagia and aspiration associated with stroke

## 3 Database acquisition and future lines

At this moment a speech database is being acquired in the protocol depicted in figure 1, one sample after deglutition of each liquid and volume. Thus, a total of 9 realizations per patient are acquired.

Figure 3 shows the acquisition scenario at Mataro's Hospital. The acquisition setup is based on a capacitor microphone Rode NT2000 (positioned at a distance of approximately 20 cm from the speaker's mouth) and external sampling card (M-AUDIO, FAST TRACK PRO Interface audio 4x4) operating at 48 kHz sampling rate, 16 bit per sample, monophonic recording. Currently we are acquiring 3 patients per week.

The signal processing approach, after database collection will be based on:

(a) Voiced/unvoiced classification and then to check the harmonic to noise ratio (HNR) on the vowels, jitter, shimmer, etc.
(b) To align the sample before and after eating using Dynamic Time Warping. The higher the distance between both realizations, the higher the probability to have deglutition problems.
(c) Some complexity measures

### 3.1 Gold Standard

Videofluoroscopy (VFS) is the gold standard to study the oral and pharyngeal mechanisms of dysphagia. VFS is a dynamic exploration that evaluates the safety and efficacy of deglutition, characterizes the alterations of deglutition in terms of videofluoroscopic signs, and helps to select and assess specific therapeutic strategies. Since the hypopharynx is full of contrast when the patient inhales after swallowing. Thereafter, VFS can determine whether aspiration is associated with impaired glossopalatal seal (predeglutitive aspiration), a delay in triggering the pharyngeal swallow or impaired deglutitive airway protection (laryngeal elevation, epiglottic descent, and closure of vocal folds during swallow response), or an ineffective pharyngeal clearance (post swallowing aspiration) [5].

___

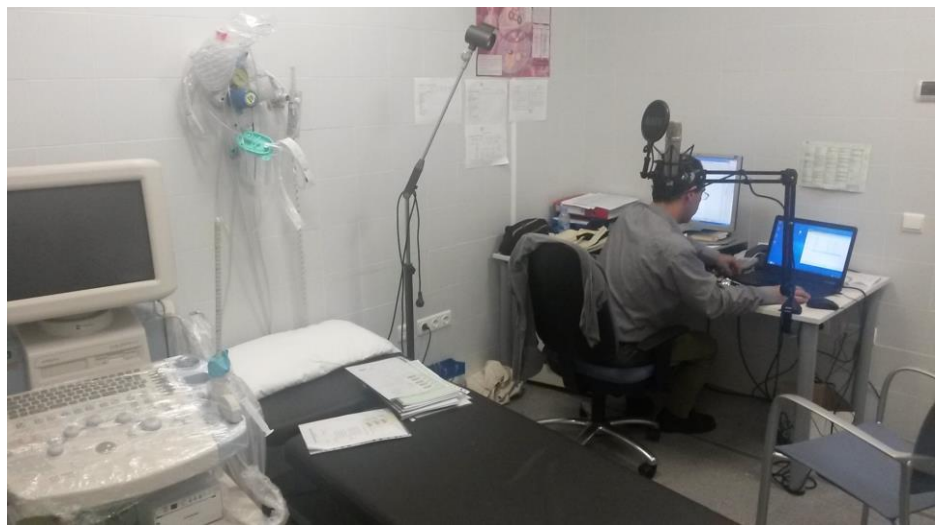

**Fig. 3.** Acquisition scenario at Mataro's Hospital

## Acknowledgement

This work has been supported by FEDER and Ministerio de ciencia e Innovación, TEC2012-38630-C04-03. The described research was performed in laboratories supported by the SIX project; the registration number CZ.1.05/2.1.00/03.0072, the operational program Research and Development for Innovation.